\documentclass[aps,prl,reprint,superscriptaddress]{revtex4-1}

\usepackage{siunitx}
\usepackage{braket}
\usepackage{graphicx}
\usepackage{amsmath}
\usepackage{amssymb}
\usepackage{comment}

\newcommand{\fig}[1]{{Fig.\ #1}}
\newcommand{\efig}[1]{{Fig.\ S#1}}
\newcommand{\cnuc}[0]{$\mathrm{^{13}C}$~}
\newcommand{\etab}[1]{{Table S#1}}

\makeatletter
\newcommand*{\balancecolsandclearpage}{
  \close@column@grid
  \cleardoublepage
  \twocolumngrid
}
\makeatother

\DeclareSIUnit\gauss{G}

\begin{document}

\title{Experimental demonstration of memory-enhanced quantum communication}

\author{M. K. Bhaskar}
\thanks{These authors contributed equally.}
\affiliation{Department of Physics, Harvard University, Cambridge, MA 02138}
\author{R. Riedinger}
\thanks{These authors contributed equally.}
\affiliation{Department of Physics, Harvard University, Cambridge, MA 02138}
\author{B. Machielse}
\thanks{These authors contributed equally.}
\affiliation{Department of Physics, Harvard University, Cambridge, MA 02138}
\author{D. S. Levonian}
\thanks{These authors contributed equally.}
\affiliation{Department of Physics, Harvard University, Cambridge, MA 02138}
\author{C. T. Nguyen}
\thanks{These authors contributed equally.}
\affiliation{Department of Physics, Harvard University, Cambridge, MA 02138}
\author{E. N. Knall}
\affiliation{John A. Paulson School of Engineering and Applied Sciences, Cambridge, MA 02138}
\author{H. Park}
\affiliation{Department of Physics, Harvard University, Cambridge, MA 02138}
\affiliation{Department of Chemistry and Chemical Biology, Harvard University, Cambridge, MA 02138, USA}
\author{D. Englund}
\affiliation{Research Laboratory of Electronics, MIT, Cambridge, MA 02139, USA}
\author{M. Lon\v{c}ar}
\affiliation{John A. Paulson School of Engineering and Applied Sciences, Cambridge, MA 02138}
\author{D. D. Sukachev}
\affiliation{Department of Physics, Harvard University, Cambridge, MA 02138}
\author{M. D. Lukin}
\thanks{lukin@physics.harvard.edu}
\affiliation{Department of Physics, Harvard University, Cambridge, MA 02138}

\begin{abstract}
The ability to communicate quantum information over long distances is of central importance  in quantum science and engineering \cite{Kimble2008}. 
For example, 
it enables secure quantum key distribution (QKD) \cite{Bennett1984,Shor2000} 
relying on fundamental physical principles that prohibit the ``cloning" of unknown quantum states \cite{Wootters1982, Dieks1982}. While QKD is already being successfully deployed \cite{Gisin2002, Boaron2018, Zhang2018, Pirandola2019a}, its range is currently limited by photon losses
and cannot be extended using straightforward measure-and-repeat strategies without compromising its unconditional security \cite{Pirandola2017}.
Alternatively, quantum repeaters \cite{Briegel1998}, which utilize intermediate quantum memory nodes and error correction techniques, can extend the range of quantum channels. However, their implementation remains an outstanding challenge \cite{Chou2007, Yuan2008, Gao2012, Reiserer2015, Hensen2015, Kalb2017}, requiring a combination of efficient and high-fidelity quantum memories, gate operations, and measurements.
Here we report the experimental realization of memory-enhanced quantum communication. 
We use
a single solid-state spin memory integrated in a nanophotonic diamond resonator \cite{Evans2018, Burek2017, Nguyen2019} 
to implement asynchronous photonic Bell-state measurements. 
This enables a four-fold increase in the secret key rate of measurement device independent (MDI)-QKD over the loss-equivalent
direct-transmission method while operating at megahertz clock rates. 
Our results represent a significant step towards
practical quantum repeaters and large-scale quantum networks \cite{Khabiboulline2018,Monroe2014}.
\end{abstract}

\maketitle

Efficient, long-lived quantum memory nodes
are expected to play an essential role in extending
the range of quantum communication \cite{Briegel1998}, as they enable asynchronous quantum logic operations, such as Bell-state measurements (BSM), between optical photons. 
For example, the BSM is crucial to MDI-QKD \cite{Lo2012, Braunstein2012}, which is a specific
implementation of quantum cryptography illustrated in \fig{1a}. 
Two remote communicating parties, Alice and Bob, try to agree on a key that is secure against potential eavesdroppers. 
They each send a randomly chosen photonic qubit $\{\ket{\pm x}, \ket{\pm y}\}$ encoded in one of two conjugate bases (X or Y) across a lossy channel to an untrusted central node (Charlie), who is asked to perform a
BSM and report the result over an authenticated
public channel. 
After a number of iterations, Alice and Bob publicly reveal their choice of bases to obtain a sifted key from the cases when they used a compatible basis.
 A provably secure key can subsequently be extracted provided the BSM error rate is low enough.
While MDI-QKD can be implemented with just linear optics and single photon detectors,
the BSM in this ``direct-transmission" approach 
is only successful when photons from Alice and Bob arrive simultaneously. Thus, when Alice and Bob are separated by a lossy fiber with a total transmission probability $p_{\textrm{A}\rightarrow\textrm{B}}\ll1$, Charlie measures photon coincidences with probability also limited by $p_{\textrm{A}\rightarrow\textrm{B}}$,  
leading to a fundamental bound \cite{Pirandola2017} on the maximum possible secret key rate of $ R_\textrm{max} = p_{\textrm{A}\rightarrow\textrm{B}}/2$ bits per channel use for an unbiased basis choice \cite{Gisin2002}.
While linear optical techniques to circumvent this bound are now being actively explored \cite{Minder2019}, they
offer only limited 
improvement
and cannot be scaled beyond a single intermediate node. 
Alternatively, this bound can be broken using a quantum memory node
at Charlie's location.
In this approach, illustrated in \fig{1b}, 
the state of Alice's photon
is efficiently stored in the heralded memory while awaiting receipt of 
Bob's
photon over the lossy channel. Once the second photon arrives,
a BSM between Alice's and Bob's qubits yields  
a secret key rate that for an ideal memory scales as $R_s \propto \sqrt{p_{\textrm{A}\rightarrow\textrm{B}}}$ \cite{Panayi2014}, potentially leading to substantial improvement over direct transmission. Beyond this specific protocol, memory-based asynchronous Bell-state measurements are central for the realization of scalable quantum repeaters \cite{Briegel1998} with multiple intermediate nodes. 

\begin{figure*}
\begin{center}
\includegraphics[width=\textwidth]{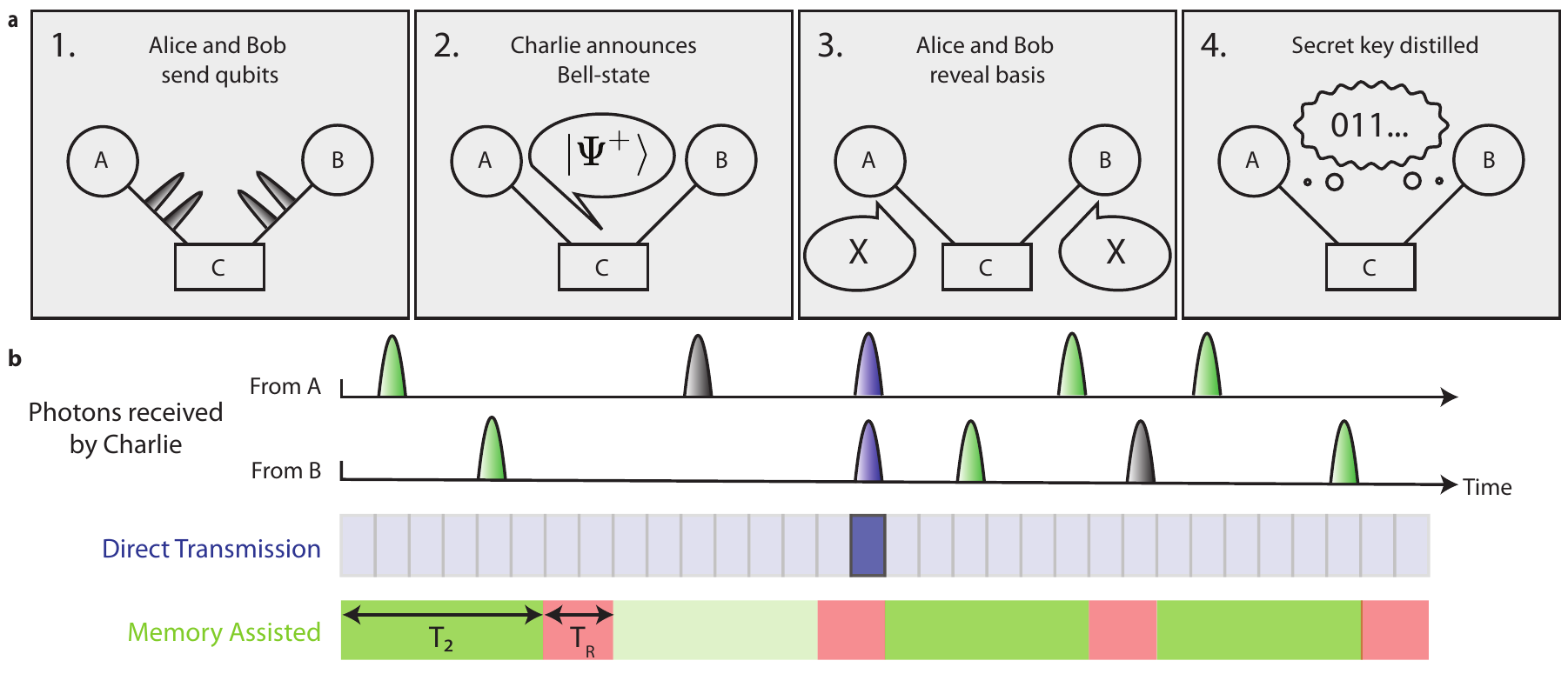}
\end{center}
\caption{{\bf Concept of memory-enhanced quantum communication.}
		{\bf a,} MDI-QKD protocol. Alice and Bob send qubits encoded in photons to a measurement device (Charlie) in between them. Charlie performs a BSM and announces the result. After verifying which rounds Alice and Bob sent qubits in compatible bases, a secure key is generated.
		{\bf b,} Illustration of memory-enhanced MDI-QKD. Photons arrive at Charlie from A and B at random times over a lossy channel, and are unlikely to arrive simultaneously (indicated in purple), leading to a low BSM success rate for direct transmission. Despite overhead time $T_R$ associated with operating a quantum memory (red), a BSM can be performed between photons that arrive at Charlie within memory coherence time $T_2$, leading to higher success rates (green). BSM successes and failures are denoted by dark and light shaded windows respectively for both approaches. 
		}
\end{figure*}

This Letter describes the operation of such a quantum memory node, enabling MDI-QKD at rates that exceed those of an ideal system based on linear optics.
Our realization 
is based on a single silicon-vacancy (SiV) color-center integrated inside a diamond nanophotonic cavity \cite{Evans2018, Burek2017, Nguyen2019} (\fig{2a}). 
Its key figure-of-merit, the cooperativity $C$ \cite{Reiserer2015}, describes the ratio of the interaction rate with individual cavity photons compared to all dissipation rates. A low mode volume ($0.5 (\lambda/n)^3$), high quality factor ($2 \times 10^4$), and nanoscale positioning of SiV centers
enable an exceptional  
$C=105\pm11$.
Cavity photons are
critically coupled to a waveguide
and adiabatically
transferred into a single-mode optical fiber \cite{Burek2017}
that is routed to superconducting nanowire single-photon detectors, yielding a full system detection efficiency of about $85\%$ \cite{SOM}.
The device is placed 
inside a dilution refrigerator, resulting in electronic spin 
quantum memory time $T_2 > $ \SI{0.2}{\milli\second} \cite{Nguyen2019}. 

The operating principle of the SiV-cavity based spin-photon interface is illustrated in \fig{2}.
Spin dependent modulation of the cavity reflection
at incident probe frequency $f_0$ (\fig{2b}) results in
the direct observation of electron spin quantum jumps (\fig{2c, inset}),
enabling nondestructive single-shot readout of the spin state (\fig{2c}) in \SI{30}{\micro\second} with fidelity $F = 0.9998^{+0.0002}_{-0.0003}$. 
Coherent control of the SiV spin qubit ($f_Q \approx $ \SI{12}{\giga\hertz}) is accomplished using microwave fields delivered 
via an on-chip gold coplanar waveguide \cite{Nguyen2019}.  
We utilize both optical readout and microwave control to perform projective feedback-based initialization of the SiV spin into the $\ket{\downarrow}$ state with a fidelity of $F = 0.998 \pm 0.001$. 
Spin-dependent cavity reflection also enables quantum logic operations between an incoming photonic time-bin qubit and the spin memory \cite{Nguyen2019, Duan2004}. We characterize this by
using the protocol illustrated in \fig{2d} to  generate the spin-photon entangled state 
$(\ket{e \uparrow} + \ket{l \downarrow})/\sqrt{2}$ conditioned on successful reflection of an incoming single photon with overall heralding efficiency $\eta = 0.423 \pm 0.004$ \cite{SOM}. Here, $\ket{e}$ and $\ket{l}$ denote the presence of a photon in an early or late time-bin separated by $\delta t = $ \SI{142}{\nano\second} respectively. 
We characterize the entangled state 
by performing measurements in the joint spin-photon ZZ and XX bases (\fig{2e}), implementing local operations on the reflected photonic qubit with a time-delay interferometer (\fig{2a}, dashed box).
By lowering the average number of photons $\braket{n}_m$ incident on the device during the SiV memory time, we reduce the possibility that an additional photon reaches the cavity without being subsequently detected, enabling high spin-photon gate fidelities for small $\braket{n}_m$ (\fig{2f}).
For $\langle n \rangle_{m} = 0.002$ we measure a lower bound on the fidelity \cite{Nguyen2019} of the spin-photon entangled state of $F \geq 0.944 \pm 0.008$, primarily limited by residual reflections from the $\ket{\downarrow}$ state. 

\begin{figure}
\begin{center}
	\includegraphics[width=\linewidth]{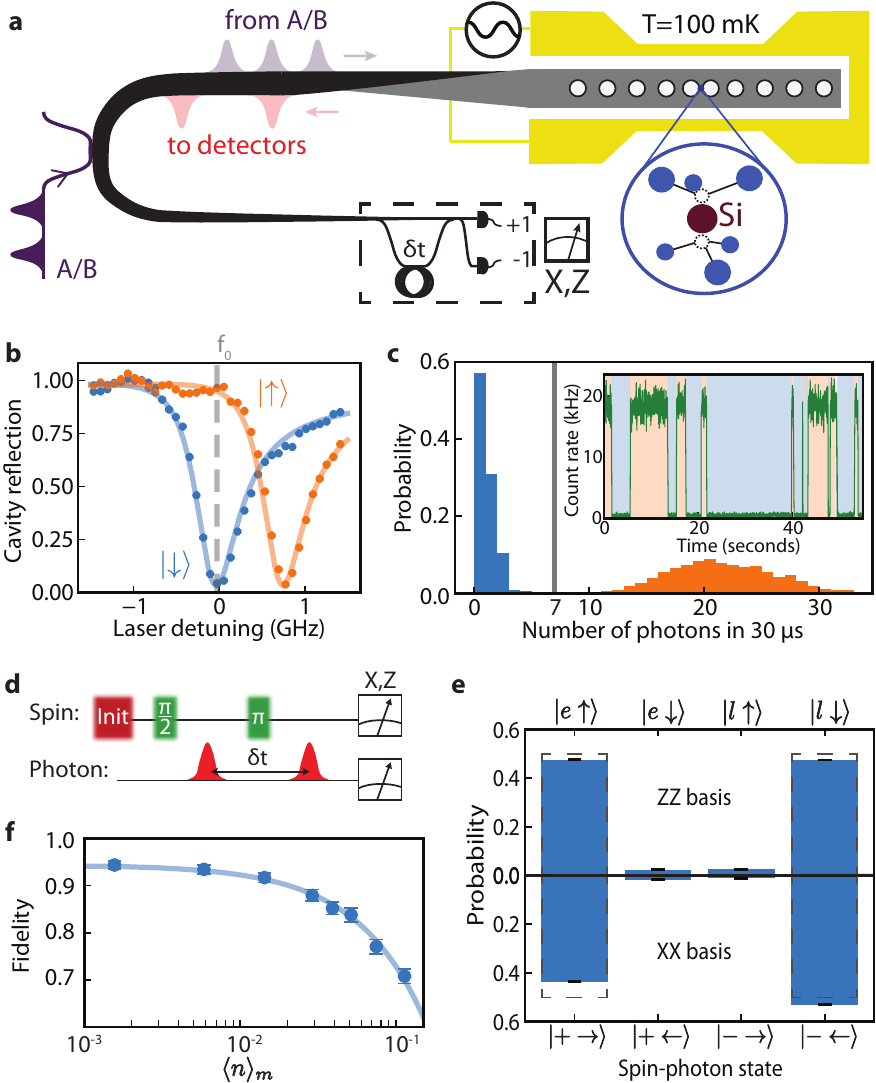}
\end{center}
\caption{{\bf Heralded spin-photon gate.}
		{\bf a,} Schematic of memory-assisted implementation of Charlie's measurement device. Weak pulses derived from a single laser simulate incoming photons from Alice and Bob (purple). Reflected photons (red) are detected in a heralding setup (dashed box).
		{\bf b,} Reflection spectrum of memory node, showing spin-dependent device reflectivity.
		{\bf c,} Histogram of detected photon numbers during a \SI{30}{\micro\second} laser pulse, enabling single-shot readout based on a threshold of $7$ photons. (Inset) Electron spin quantum jumps under weak illumination.
		{\bf d,} Schematic of spin-photon quantum logic operation used to generate and verify spin-photon entangled state.
		{\bf e,} Characterization of resulting spin-photon correlations in the ZZ and XX bases. Dashed bars show ideal values.
		{\bf f,} Measured spin-photon entanglement fidelity as a function of $\langle n \rangle_{m}$, the average incident photon number during each initialization of the memory.
		}
\end{figure}

This spin-photon logic gate can be directly used to herald the storage of an incoming photonic qubit by interferometrically measuring the reflected photon in the X basis \cite{Nguyen2019}. To implement memory-assisted MDI-QKD, we extend this protocol  to accommodate a total of $N$ photonic qubit time-bins within a single initialization of the memory (\fig{3a}). Each individual time-bin qubit is encoded in the relative amplitudes and phases of a pair of neighboring pulses separated by $\delta t$. 
Detection of a reflected photon 
heralds the arrival of the photonic qubit formed by the two interfering pulses without revealing its state \cite{Nguyen2019}.
Two such heralding events, combined with subsequent spin-state readout in the $X$ basis, constitute a successful BSM on the incident photons.
This can be understood without loss of generality 
by restricting 
input 
photonic states to be encoded in the relative phase $\phi$ between 
neighboring pulses
with equal amplitude: $(\ket{e} + e^{i \phi} \ket{l})/\sqrt{2}$ (\fig{3b}). 
Detection of the first reflected photon in the X basis teleports its quantum state onto the spin, resulting in the state $(\ket{\uparrow} + m_1 e^{i \phi_1} \ket{\downarrow})/\sqrt{2}$, where $m_1= \pm 1$ depending on which detector registers the photon \cite{Nguyen2019}.
Detection of a second photon at a later time within the electron spin $T_2$ results in the spin state $(\ket{\uparrow} + m_1 m_2 e^{i(\phi_1 + \phi_2)} \ket{\downarrow})/\sqrt{2}$. The phase of this spin state depends only on the sum of the incoming phases and the product of their detection outcomes, but not the individual phases themselves. 
As a result, if the photons were sent with phases that meet the condition $\phi_1 + \phi_2 \in \{0, \pi\}$, a final measurement of the spin in the $X$ basis $(m_3 = \pm 1)$ completes an asynchronous photon-photon BSM, distinguishing two of the four Bell-states 
based on the total parity $m_1 m_2 m_3 = \pm1$ \cite{SOM}.

\begin{figure*}
\begin{center}
	\includegraphics[width=\textwidth]{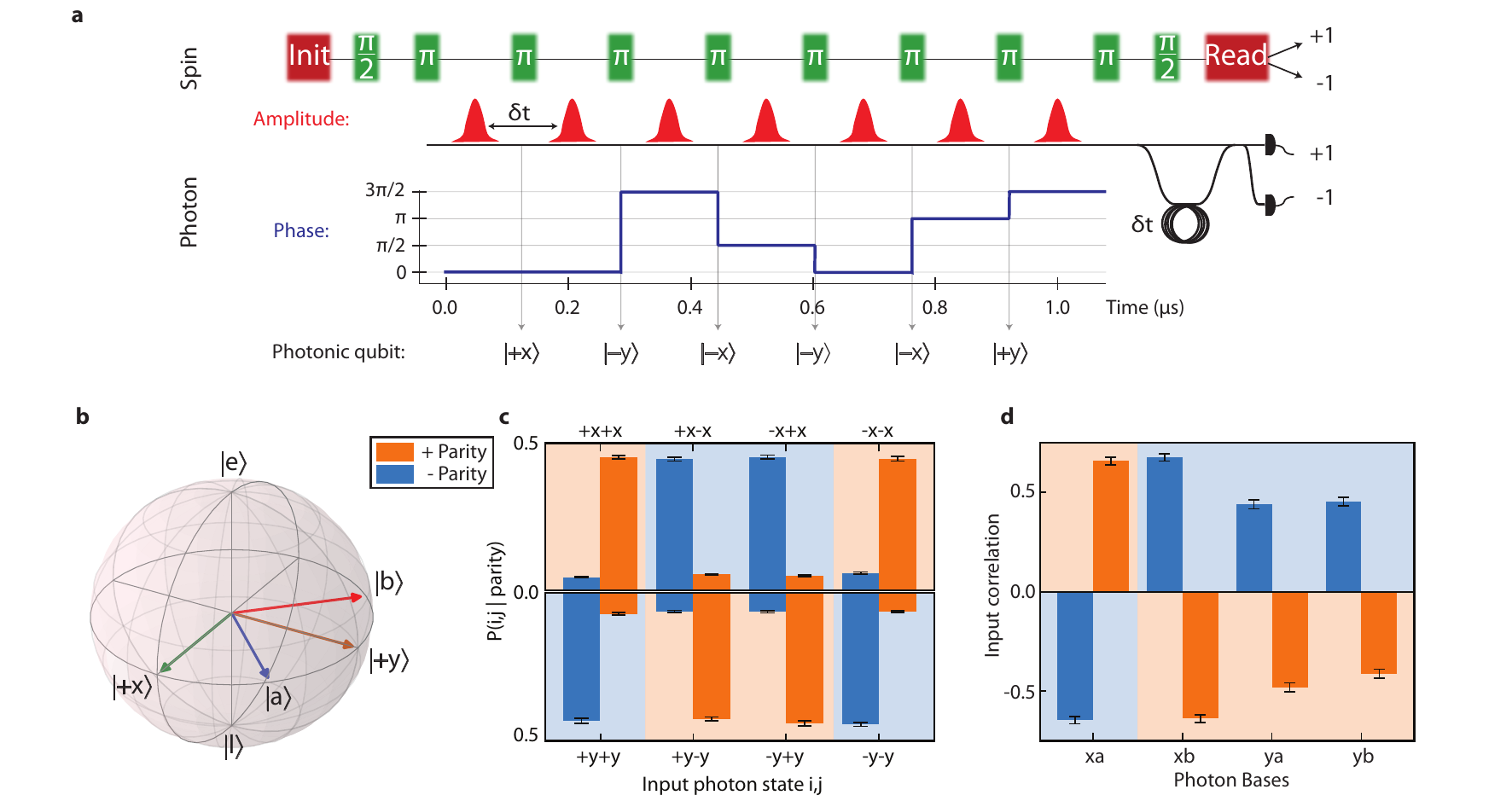}
\end{center}
\caption{{\bf Asynchronous Bell-state measurements using quantum memory.}
		{\bf a,} Example sequence with $N=6$ photonic qubits sent in a single memory time. Microwave $\pi$ pulses (green) are interleaved with incoming optical pulses. Photons have fixed amplitude (red) and qubits are defined by the relative phases between subsequent pulses (blue).  
		{\bf b,} Bloch sphere representation of input photonic time-bin qubits used for characterization. 
		{\bf c,} Characterization of asynchronous BSM. Conditional probabilities for Alice and Bob to have sent input states $(i, j)$ given a particular parity outcome for input states in the $X$ (top) and $Y$ (bottom) bases. 
		{\bf d,} Bell test using the CHSH inequality. Conditioned on the BSM outcome, the average correlation between input photons is plotted for each pair of bases used \cite{SOM}. 
		Shaded backgrounds denote the expected parity.
		}
\end{figure*}

This approach can be directly applied to generate a secure key within the MDI-QKD protocol illustrated in \fig{1a}.
We analyze the system performance
by characterizing the overall quantum-bit error rate (QBER) \cite{Gisin2002, Lo2012} for $N = 124$ photonic qubits per memory initialization.
We use several random bit strings of incoming photons from $\{\ket{\pm x}, \ket{\pm y}\}$ and observe strong correlations between the resulting BSM outcome and the initial combination of input qubits for both bases (\fig{3c}).
Using this method, we estimate the average QBER to be 
$E = 0.116\pm0.002$
for all combinations of random bit strings measured,
significantly below the limit of $E_i=0.146$ providing security against individual attacks \cite{Gisin2002}. 
This value is affected by technical imperfections in the preparation of random strings of photonic qubits. 
We find specific periodic patterns of photonic qubits to be less prone to these effects, resulting
in a QBER as low as $E = 0.097\pm0.006$, which falls within the threshold for unconditional security of $E_u=0.110$ \cite{Shor2000} with a confidence level of $0.986$ \cite{SOM}. 
We further verify security by testing the Bell-CHSH inequality \cite{Hensen2015} using input states from four different bases, each separated by an angle of $45^\circ$ \cite{SOM}. 
We find that the correlations between input photons (\fig{3d}) violate the Bell-CHSH inequality $S_\pm \leq 2$, observing $S_+ = 2.21 \pm 0.04$ and $S_- = 2.19 \pm 0.04$ for positive and negative BSM parity results respectively. This result demonstrates that this device can be used for fundamentally secure quantum communication \cite{Gisin2002}. 

Finally, we benchmark the performance of memory-assisted QKD. For each experiment, 
we model an effective channel loss by considering the mean photon number $\braket{n}_p$ incident on the device per photonic qubit. Assuming that 
Alice and Bob emit roughly one photon per qubit, this yields an effective channel 
transmission probability $p_{A\rightarrow B} = \braket{n}_p^2$, resulting in
the maximal secret key rate $R_\textrm{max}$ per channel use for direct transmission MDI-QKD \cite{Lo2012}, given by the red line in \fig{4}. We emphasize that this is a theoretical upper bound on linear optics based MDI-QKD, assuming 
ideal sources and detectors and balanced basis choices.
The measured sifted key rates of the memory-based device are plotted as open circles in \fig{4}. 
Due to the high overall heralding efficiency and the large number of photonic qubits per memory time (up to $N = 504$), the memory-assisted sifted key rate exceeds the capability of direct-transmission MDI-QKD by a factor of 
$78.4 \pm 0.7$ at an effective channel loss of about $88$ \si{dB}.

In practice, errors introduced by the quantum memory node could leak information to the environment, reducing the security of the sifted key \cite{Shor2000}. The fraction of secure bits $r_s$ that can be extracted from a sifted key with finite QBER using conventional error correction and privacy amplification techniques 
rapidly diminishes \cite{Gisin2002} as the QBER approaches $E_i = 0.147$. 
For each value of the effective channel loss, we estimate the QBER and use it to compute $r_s$, enabling extraction of distilled secure key rates $R_\textrm{S}$, plotted in black in \fig{4}. 
Even after error-correction, we find that the memory-assisted secret key rate outperforms the ideal limit for the corresponding direct-transmission implementation of MDI-QKD 
by a factor of up to $R_\textrm{S}/R_\textrm{max} = 4.1\pm 0.5$ 
($\pm0.1$ systematic uncertainty, for $N=124$). 
We further find that this rate also exceeds the fundamental bound on repeaterless communication \cite{Pirandola2017} $R_\textrm{S}\leq 1.44p_{\textrm{A}\rightarrow\textrm{B}}$ with a statistical confidence level of 
$99.2\%$ ($\hspace{0pt}^{+0.2\%}_{-0.3\%}$ systematic uncertainty \cite{SOM}).
Despite experimental overhead time associated with operating the quantum memory node ($T_R$ in \fig{1b}), the performance of the memory assisted BSM (for $N = 248$) enables MDI-QKD that is competitive with an ideal unassisted system running at a \SI{4}{\mega\hertz} average clock rate \cite{SOM}.

\begin{figure}
\begin{center}
	\includegraphics[width=\linewidth]{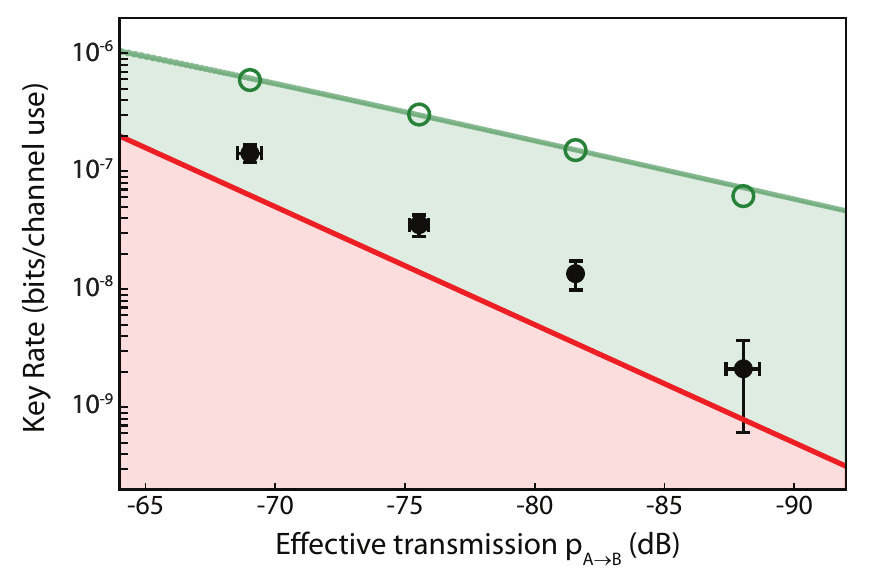}
\end{center}
\caption{
	{\bf Performance of memory-assisted quantum communication.} Log-log plot of key rate in bits per channel use versus effective channel transmission 
	($p_{\textrm{A}\rightarrow\textrm{B}}=\braket{n}_p^2$, where $\braket{n}_p$ is the average number of photons incident on the measurement device per photonic qubit).
	Red line: theoretical maximum for equivalent direct transmission MDI-QKD experiment. Green open circles: experimentally measured sifted key rate (green line is the expected rate). 
	To ensure optimal operation of the memory, $\braket{n}_m = \braket{n}_p N \approx 0.02$ is kept constant \cite{SOM}. From left to right, points correspond to $N = \{60, 124, 248, 504\}$. Black filled circles: secure key rates $R_\textrm{S}$ using memory device. 
	Vertical error bars are given by the $68\%$ confidence interval and horizontal error bars represent the standard deviation of the systematic power fluctuations.
}
\end{figure}

These experiments demonstrate the viability of memory-enhanced quantum communication and represent 
a crucial step towards realizing functional quantum repeaters. 
Several important technical improvements will be necessary to 
apply
this advance for practical quantum communication. First, this protocol must be implemented using truly independent, distant communicating parties. Additionally, frequency conversion from telecommunications wavelengths, as well as low-loss optical elements used for routing photons to and from the memory node, will need to be incorporated. 
Finally, rapid generation of provably secure keys will require implementation of decoy-state protocols \cite{Lo2005a}, biased bases \cite{Lo2005}, and finite-key analyses \cite{Curty2014}, all compatible with the present approach. 
With these improvements, our approach
is well-suited for deployment in real-world settings. 
It does not require phase stabilization of long-distance links and operates 
efficiently in the relevant regime of $p_{A\rightarrow B} \approx $ \SI{70}{dB}, 
corresponding to about \SI{350}{\kilo\meter} of telecommunications fiber.
Additionally, a single device can be used at the center of a star network topology \cite{Biham1996}, enabling quantum communication between several parties beyond the metropolitan scale.
Furthermore, the present approach can be extended along several directions. The use of long-lived \cnuc nuclear spin qubits could eliminate the need to operate at low total $\braket{n}_m$ and would provide longer storage times, potentially enabling hundred-fold enhancement of
BSM success rates \cite{Kalb2017, Nguyen2019}.
Recently implemented strain-tuning capabilities \cite{Machielse2019} should allow for operation of many quantum nodes at a common network frequency.
Unlike linear-optics based alternatives \cite{Minder2019}, the approach presented here can be extended to implement the full repeater protocol, enabling a polynomial scaling of the communication rate with distance \cite{Briegel1998}.
Finally, the demonstrated multi-photon gate operations 
can also be adapted to engineer large cluster-states of entangled photons\cite{Raussendorf2001}, which can be utilized for 
rapid quantum communication \cite{Borregaard2019}. Implementation of these techniques could enable the realization and applications of scalable quantum networks \cite{Kimble2008} beyond QKD, ranging from non-local quantum metrology \cite{Khabiboulline2018} to modular quantum computing
architectures \cite{Monroe2014}.  

\section{Acknowledgments}
\begin{acknowledgments}
We thank Pavel Stroganov, Kristiaan de Greve, Johannes Borregaard, Eric Bersin, Benjamin Dixon, and Neil Sinclair for discussions, Vikas Anant from PhotonSpot for providing SNSPDs, and Jim MacArthur for assistance with electronics. This work was supported by the NSF, CUA, DoD/ARO DURIP, AFOSR MURI, ONR MURI, ARL, and a Vannevar Bush Faculty Fellowship. Devices were fabricated at Harvard CNS, NSF award no. 1541959. M. K. B. and D. S. L. acknowledge support from an NDSEG Fellowship. R. R. acknowledges support from the Alexander von Humbolt Foundation. B. M. and E. N. K. acknowledge support from an NSF GRFP.
\end{acknowledgments}

\bibliography{Half-repeater}

\balancecolsandclearpage
\widetext
\begin{center}
\textbf{\large Supplementary Material}
\end{center}

\twocolumngrid
\setcounter{figure}{0}
\renewcommand{\thefigure}{S\arabic{figure}}
\renewcommand{\thetable}{S\arabic{table}}

\section{Experimental setup} 

Experimental setup and device fabrication \cite{Burek2014, Burek2017, Machielse2019, Atikian2017} for
millikelvin nanophotonic cavity QED experiments with SiV centers are thoroughly 
described in a separate publication \cite{Nguyen2019a}. 
We perform all measurements in a dilution refrigerator (DR, BlueFors BF-LD250) with a base temperature of 20\,mK.
The DR is equipped with a superconducting vector magnet (American Magnets Inc. 6-1-1\,T), a home-built free-space wide-field microscope with a cryogenic objective (Attocube LT-APO-VISIR), piezo positioners (Attocube ANPx101 and ANPx311 series), and fiber and MW feedthroughs.
Tuning of the nanocavity resonance is performed using a gas condensation technique \cite{Evans2018}.
The SiV-cavity system is optically interrogated through the fiber network without any free-space optics \cite{Nguyen2019}.
The operating temperature of the memory node during the BSM measurements was 100-300\,mK.

\subsection{Experimental implementation of asynchronous BSM}

An asynchronous BSM (\fig{3a}) relies on (1) precise timing of the arrival of optical pulses (corresponding to photonic qubits \cite{DeRiedmatten2004,Sasaki2014} from Alice and Bob) with microwave control pulses on the quantum memory and (2) interferometrically stable rotations on reflected time-bin qubits for successful heralding. In order to accomplish (1), all equipment used for generation of microwave and optical fields is synchronized by a single device (National Instriuments HSDIO, \efig{1a}) with programming described in \etab{1-2}. 

In order to accomplish (2), we use a single, narrow linewidth ($< $ \SI{50}{\kilo\hertz}) 
Ti:Sapphire laser (M Squared SolsTiS-2000-PSX-XF, \efig{1b}) both for generating 
photonic qubits and locking the time-delay interferometer (TDI) used to herald their arrival. 
In the experiment, photonic qubits are reflected from the device, sent into the TDI, and detected on superconducting nanowire single photon detectors (SNSPD, Photon Spot). All detected photons are processed digitally on a field-programmable gate array (FPGA, \efig{1a}), and the arrival times of these heralding signals are recorded on a time-tagger (TT, \efig{1a}), and constitute one bit of information of the BSM ($m_1$ or $m_2$). At the end of the experiment, a \SI{30}{\micro\second} pulse from the readout path is reflected off the device, and photons are counted in order to determine the spin state ($m_3$) depending on the threshold shown in \fig{2c}.

\def\arraystretch{1.5}
\begin{table*}
	\begin{center}
		\begin{tabular}{ l | c | c | c}
			\hline
			Step & Process & Duration & Proceed to \\
			\hline
			1 & Lock time-delay interferometer & \SI{200}{\milli\second} & 2 \\
			2 & Readout SiV & \SI{30}{\micro\second} & If status LOW: 4, else: 3 \\
			3 & Apply microwave $\pi$ pulse & \SI{32}{\nano\second} & 2 \\
			4 & Run main experiment script & $\sim$ \SI{200}{\milli\second} & 1 \\
			\hline
		\end{tabular}
		\caption{{\bf High-level experimental sequence}. This sequence is programmed into the HSDIO and uses feedback from the status trigger sent from the FPGA (see \efig{1a}). Main experimental sequence is described in \etab{2}.
		External software is also used to monitor the status trigger. If it is HI for $\gtrsim $ \SI{2}{\second}, the software activates an automatic re-lock procedure which compensates for spectral diffusion and ionization of the SiV center.
		}
	\end{center}
\end{table*}

\begin{table*}
	\begin{center}
		\begin{tabular}{ l | c | c | c}
			\hline
			Step & Process & Duration & Proceed to \\
			\hline
			1 & Run sequence in \fig{3a} for a given $N$ & $10-20$ \si{\micro\second}& 2 \\
			2 & Readout SiV + report readout to TT & \SI{30}{\micro\second} & If status LOW: 1, else: 3 \\
			3 & Apply microwave $\pi$ pulse & \SI{32}{\nano\second} & 4 \\
			4 & Readout SiV & \SI{30}{\micro\second} & If status LOW: 3, else: 1 \\
			\hline
		\end{tabular}
		\caption{{\bf Main experimental sequence for memory-enhanced quantum communication}. This script is followed until step $1$ is run a total of $4000$ times, and then terminates and returns to step 1 of \etab{1}. The longest step is the readout step, which is limited by the fact that we operate at a photon detection rate of $\sim$ \SI{1}{\mega\hertz} to avoid saturation of the SNSPDs.}
	\end{center}
\end{table*}

To minimize thermal drift of the TDI, it is mounted to a thermally weighted aluminum breadboard, placed in a polyurethane foam-lined and sand filled briefcase, and secured with glue to ensure passive stability on the minute timescale.
We halt the experiment and actively lock the interferometer to the sensitive Y-quadrature every $\sim$ \SI{200}{\milli\second} by changing the length of the roughly \SI{28}{\meter} long (\SI{142}{\nano\second}) delay line with a cylindrical piezo. 
In order to use the TDI for X-measurements of the reflected qubits, we apply a frequency shift of \SI{1.8}{\mega\hertz} using the qubit AOM, which is $1/4$ of the free-spectral range of the TDI.
Since the nanophotonic cavity, the TDI, and the SNSPDs are all polarization sensitive, we use various fiber-based polarization controllers (\efig{1b}).
All fibers in the network are covered with aluminum foil to prevent thermal polarization drifts.
This results in an interference visibility of the TDI of $> 99\%$ that is stable for several days without any intervention with lab temperature and humidity variations of $\pm 1^\circ$ C and $\pm 5\%$ respectively.

In order to achieve high-fidelity operations we have to ensure that the laser frequency (which is not locked) is resonant with the SiV frequency $f_0$ (which is subject to the spectral diffusion \cite{Nguyen2019a}).
To do that we implement a so-called preselection procedure, described in \etab{1-2} and \efig{1a}.
First, the SiV spin state is initialized by performing a projective measurement and applying microwave feedback.
During each projective readout, the reflected counts are compared with two thresholds: a ``readout" threshold of $7$ photons (used only to record $m_3$), and a ``status" threshold of $3$ photons. 
The status trigger is used to prevent the experiment from running in cases when the laser is no longer on resonance with $f_0$, or if the SiV has ionized to an optically inactive charge state.
The duty cycle of the status trigger is externally monitored and is used to temporarily abort the experiment and run an automated re-lock procedure that locates and sets the laser to the new frequency $f_0$, reinitailizing the SiV charge state with a \SI{520}{\nano\meter} laser pulse if nececssary. 
This protocol enables fully automated operation at high fidelities (low QBER) for several days without human intervention.

\subsection{Calibration of fiber network}

\begin{figure*}
\begin{center}
	\includegraphics[width=\textwidth]{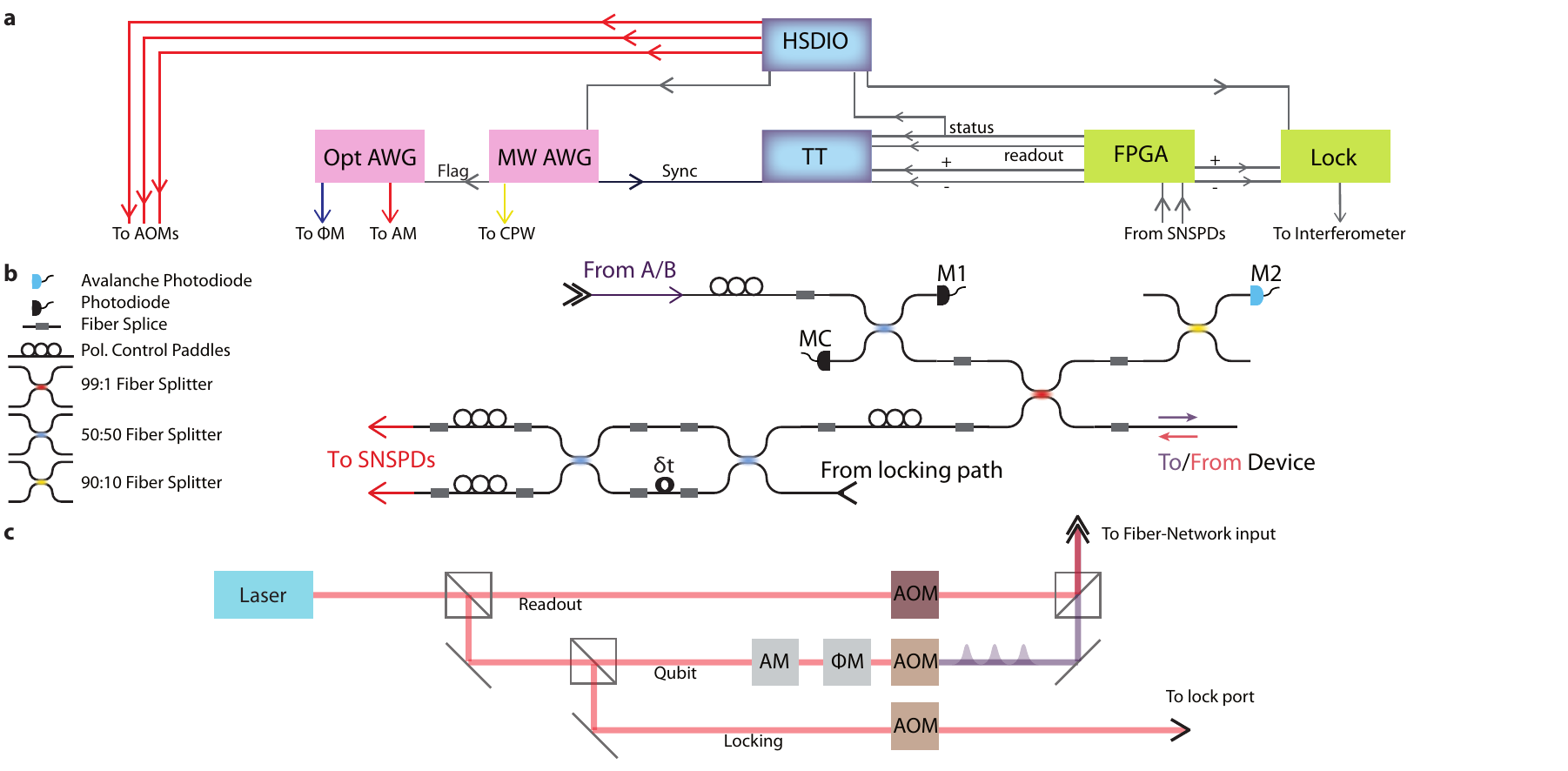}
\end{center}
\caption{{\bf Experimental schematic.}
		{\bf a,} Control flow of experiment. Opt (MW) AWG is a Tektronix AWG7122B \SI{5}{GS/s} (Tektronix AWG70001a \SI{50}{GS/s}) arbitrary waveform generator used to generate photonic qubits (microwave control signals).
		All signals are recorded on a time-tagger (TT, PicoQuant HydraHarp 400). 
		{\bf b,} Fiber network used to deliver photons to and collect photons from the memory device, including elements for polarization control and diagnostic measurements of coupling efficiencies. 
		{\bf c,} Preparation of optical fields. The desired phase relation between lock and qubit paths is ensured by modulating AOMs using phase-locked RF sources with a precise \SI{1.8}{\mega\hertz} frequency shift between them.
		}
\end{figure*}

The schematic of the fiber-network used to deliver optical pulses to and collect reflected photons from the nanophotonic memory device is shown in \efig{1b}. 
Photons are routed through the lossy ($1\%$) port of a 99:1 fiber beamsplitter (FBS) to the 
nanophotonic device.
We note that for practical implementation of memory-assisted quantum communication, an efficient optical switch or circulator should be used instead.
In this experiment, since we focus on benchmarking the performance of the memory device itself, 
the loss introduced by this beamsplitter 
is incorporated into the estimated channel loss. Reflected photons are collected and routed back through the efficient ($99\%$) port of the FBS and are sent to the TDI in the heralding setup.

The outputs of the TDI are sent
back into the dilution refrigerator and directly coupled to superconducting nanowire single-photon detectors (SNSPDs, PhotonSpot), which are mounted at the 1K stage and are coated with dielectrics to optimize detection efficiency exactly at \SI{737}{\nano\meter}. 
To estimate the quantum efficiency (QE) of the detectors we compare the performance of the SNSPDs to the specifications of calibrated conventional avalanche photodiodes single-photon counters (Laser Components COUNT-10C-FC). The estimated QEs of the SNSPDs with this method are as close to unity as we can verify. Additionally, we measure $< 1\%$ reflection from the fiber-SNSPD interface, which typically is the dominant contribution to the reduction of QE in these devices.
Thus we assume the lower bound of the QE of the SNSPDs to be $\eta_{\mathrm{QE}} = 0.99$ for the rest of this section.
Of course, this estimation is subject to additional systematic errors. However, the actual QE of these detectors would be a common factor (and thus drop out) in a comparison between any two physical quantum communication systems.

The total heralding efficiency $\eta$ of the memory node is an important parameter since it directly affects the performance of the BSM for quantum communication experiments.
Here we use 2 different approaches to estimate the overall heralding efficiency $\eta$.
We first measure the most dominant loss, which arises from the average reflectivity of the critically coupled nanophotonic cavity (\fig{2b}). 
While the $\ket{\uparrow}$ state is highly reflecting ($94.4\%$), the $\ket{\downarrow}$ state reflects only $4.1\%$ of incident photons, leading to an average device reflectivity of $\eta_{sp} = 0.493$. 

In method (1), we compare the input power photodiode M1 with that of photodiode MC. This estimates a lower-bound on the tapered-fiber diamond waveguide coupling efficiency of $\eta_c = 0.930 \pm 0.017$. This error bar arises from uncertainty due to photodiode noise and does not include systematic photodiode calibration uncertainty. 
However, we note that if the tapered fiber is replaced by a silver-coated fiber-based retroreflector, this calibration technique extracts a coupling efficiency of $\eta_c^{cal} \approx 0.98$, which is consistent with the expected reflectivity from such a retroreflector.
We independently calibrate the efficiency through the 99:1 fiber beamsplitter and the TDI to be $\eta_f = 0.934$. This gives us our first estimate on the overall heralding efficiency $\eta = \eta_{sp} \eta_c \eta_f \eta_{\mathrm{QE}} = 0.425 \pm 0.008$. 

In method (2), during the experiment we compare the reflected counts from the highly-reflecting ($\ket{\uparrow}$) spin-state measured on the SNSPDs with the counts on an avalanche photodiode single photon counting module (M2 in \efig{1b}) which has a calibrated efficiency of $\approx 0.7$ relative to the SNSPDs. From this measurement, we estimate an overall efficiency of fiber-diamond coupling, as well as transmission through all relevant splices and beamsplitters of $\eta_c \eta_f = 0.864 \pm 0.010$. This error bar arises from shot noise on the single photon detectors. Overall, this gives us a consistent estimate of $\eta = \eta_{sp} \eta_c \eta_f \eta_{\mathrm{QE}} = 0.422 \pm 0.005$.

For values cited in the main text and data points presented in the figures, we use 
an average value of the heralding efficiency inferred from the two calibration techniques: $\eta = 0.423\pm0.004$.
Methods (1) and (2), which each have independent systematic uncertainties associated with imperfect photodetector calibrations, are consistent to within a small residual systematic uncertainty, which is noted in the text where appropriate.
We note that this heralding efficiency is consistent with the scaling of spin decoherence with the number of photons at the cavity $\braket{n}_{m}$. An example of this effect is shown in the red point in \efig{3e}.

\section{Characterization of the nanophotonic quantum memory.}

\begin{figure}
\begin{center}
	\includegraphics[width=\linewidth]{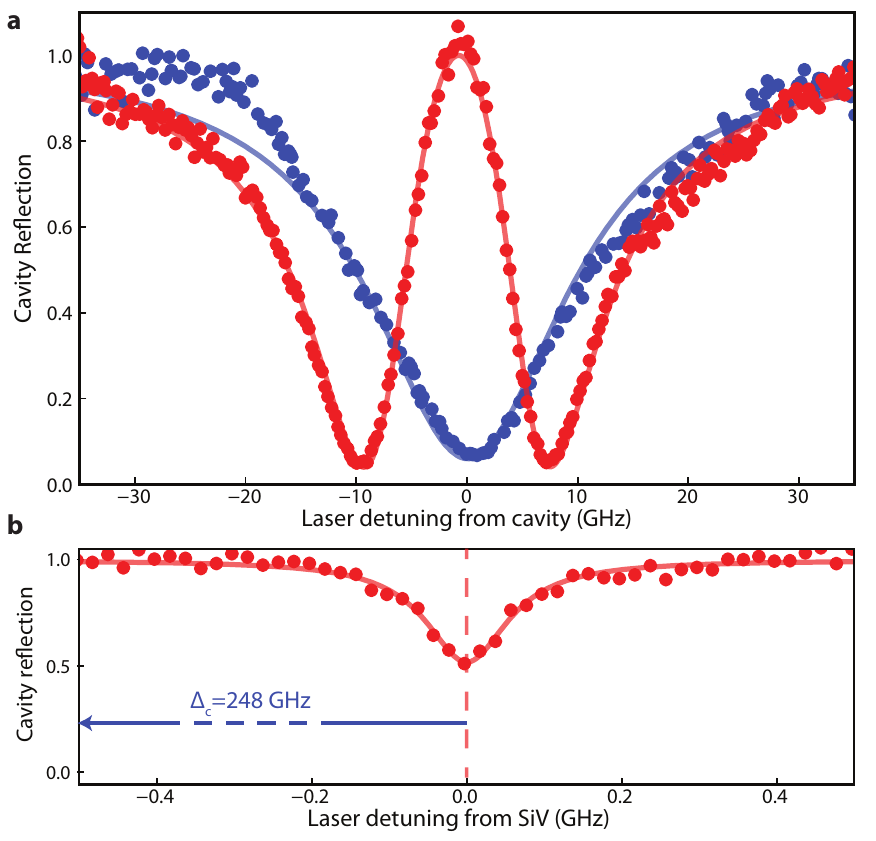}
\end{center}
\caption{{\bf Characterization of device cooperativity.}
		{\bf a,} Cavity reflection spectrum far-detuned (blue) and on resonance (red) with SiV center. Blue solid line is a fit to a Lorentzian, enabling extraction of linewidth $\kappa = $ \SI{21.8}{\giga\hertz}. Red solid line is a fit to a model used to determine the single-photon Rabi frequency $g = 8.38 \pm 0.05$ \si{\giga\hertz} and shows the onset of a normal mode splitting.
		{\bf b,} Measurement of SiV linewidth far detuned ($\Delta_c = $ \SI{248}{\giga\hertz}) from cavity resonance. Red solid line is a fit to a Lorentzian, enabling extraction of natural linewidth $\gamma = $ \SI{0.123}{\giga\hertz}.
		}
\end{figure}

\begin{figure*}
\begin{center}
	\includegraphics[width=\textwidth]{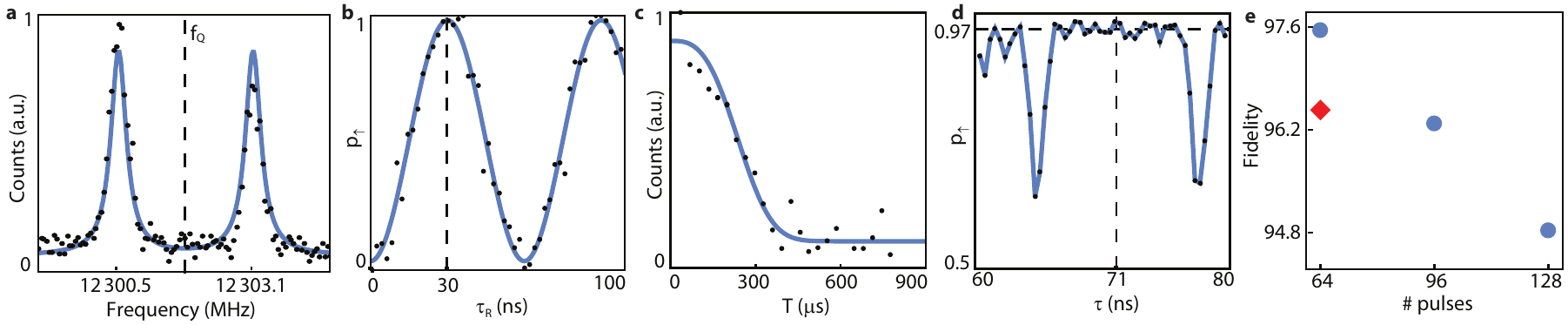}
\end{center}
\caption{{\bf Microwave characterization of spin-coherence properties.}
		{\bf a,} 
		ODMR spectrum of the qubit transition at $\sim$ \SI{12}{\giga\hertz} split by coupling to a nearby \cnuc.
		{\bf b,} Rabi oscillations showing $\pi$ time of \SI{30}{\nano\second}. A $\pi$ time of \SI{32}{\nano\second} is used for experiments in the main text.
		{\bf c,} XY8-1 dynamical decoupling signal (unnormalized) as a function of total time $T$, showing coherence lasting on the several hundred \si{\micro\second} timescale.
		{\bf d,} XY8-8 dynamical decoupling signal (normalized) revealing region of high fidelity at relevant value of $2 \tau = $ \SI{142}{\nano\second}.
		{\bf e,} Fidelity of spin state after dynamical decoupling sequence with varying number of $\pi$ pulses ($N_\pi$), blue points. Red point (diamond) is under illumination with $\braket{n}_m = 0.02$.
		}
\end{figure*}

A spectrum of the SiV-cavity system at large detuning (248\,GHz) allows us to measure the cavity linewidth $\kappa=21.6 \pm 1.3$\,GHz, (\efig{2a}, blue curve) and natural SiV linewidth $\gamma = 0.123 \pm 0.010$\,GHz (\efig{2a}, red curve).
We find spectral diffusion of the SiV optical frequency to be much smaller than $\gamma$ on minute timescales with an excitation photon flux of less than \SI{1}{\mega\hertz}.
Next, we estimate the single-photon Rabi frequency, $g$, using the cavity reflection spectrum for zero atom-cavity detuning, shown in red in \efig{2a}. 
For a resonant atom-cavity system probed in reflection from a single port with cavity-waveguide coupling $\kappa_{wg}$, the cavity reflection coefficient \cite{Reiserer2015} as a function of probe detuning $\Delta_c$ is given by 
\begin{equation}
r(\Delta_c) = \frac{i\Delta_c + \frac{g^2}{i \Delta_c + \frac{\gamma}{2}} - \kappa_{wg} + \frac{\kappa}{2}} {i\Delta_c + \frac{g^2}{i \Delta_c + \frac{\gamma}{2}} + \frac{\kappa}{2}}.
\end{equation}
By fitting $|r(\Delta_c)|^2$ using known values of $\kappa$ and $\gamma$, we obtain the solid red curve in \efig{2a} which corresponds to a single-photon Rabi frequency $g = 8.38 \pm 0.05$ \si{\giga\hertz}, yielding the estimated cooperativity $C =  \frac{4g^2}{\kappa \gamma}=105 \pm 11$.

We use resonant MW pulses delivered via an on-chip coplanar waveguide (CWG) to coherently control the quantum memory \cite{Nguyen2019,Nguyen2019a}.
First, we measure the spectrum of the spin-qubit transition by applying a weak, \SI{10}{\micro\second}-long microwave pulse of variable frequency, observing the optically-detected magnetic resonance (ODMR) spectrum presented in \efig{3a}. We note that the spin-qubit transition is split by the presence of a nearby \cnuc. While coherent control techniques can be employed to utilize the \cnuc~as an additional qubit \cite{Nguyen2019,Nguyen2019a}, we do not control or initialize it in this experiment. Instead, we drive the electron spin with strong microwave pulses at a frequency $f_Q$ such that both \cnuc-state-specific transitions are addressed equally. This also mitigates slow spectral diffusion of the microwave transition \cite{Nguyen2019a} of $\sim$ \SI{100}{\kilo\hertz}. 

After fixing the MW frequency at $f_Q$ we vary the length of this drive pulse ($\tau_R$ in \efig{3b}) and observe full-contrast Rabi oscillations. 
We choose a $\pi$ time of \SI{32}{\nano\second} in the experiments in the main text, which is a compromise of two factors: (1) it is sufficiently fast such that we can temporally multiplex between $2$ and $4$ time-bin qubits around each microwave $\pi$ pulse and (2) it is sufficiently weak to minimize heating related effects from high microwave currents in  resistive gold CWG.

With known $\pi$ time we measure the coherence time of the SiV spin qubit under an XY8-1 dynamical decoupling sequence to exceed \SI{200}{\micro\second} (\efig{3c}).
In the main experiment we use decoupling sequences with more $\pi$ pulses.
As an example, \efig{3d} shows the population in the $\ket{\uparrow}$ state after XY8-8 decoupling sequence (total $N_\pi = 64$ $\pi$ pulses) as a function of $\tau$, half of the inter-pulse spacing.
For BSM experiments, this inter-pulse spacing, $2\tau$, is fixed and is matched to the time-bin interval $\delta t$.
While at some times (e.g. $\tau = $ \SI{64.5}{\nano\second}) there is a loss of coherence due to entanglement with the nearby \cnuc, at $2\tau = $ \SI{142}{\nano\second} we are decoupled from this \cnuc~and can maintain a high degree of spin coherence.
Thus we chose the time-bin spacing to be 142\,ns.
The spin coherence at $2\tau = $ \SI{142}{\nano\second} is plotted as a function $N_\pi$ in \efig{3d}, and decreases for large $N_\pi$, primarily due to heating related effects \cite{Nguyen2019}.

\section{Theoretical description of asynchronous Bell state measurement} Due to the critical coupling of the nanocavity, the memory node only reflects photons when the SiV spin is in the state $\ket \uparrow$. The resulting correlations between the spin and the reflected photons can still be used to realize a BSM between two asynchronously arriving photonic time-bin qubits using an adaptation of the well known proposal of Duan and Kimble \cite{Duan2004} for entangling a pair of photons incident on an atom-cavity system. 
As a result of the critical coupling, we only have access to two of the four Bell states at any time, with the inaccessible Bell states corresponding to photons being transmitted through the cavity (and thus lost from the detection path). 
Depending on whether there was an 
even or odd number of $\pi$-pulses on the spin between the arrival of the two heralded photons, we distinguish either the $\{\ket{\Phi_\pm}\}$ or $\{\ket{\Psi_\pm}\}$ states (defined below). For the sake of simplicity, we first describe the BSM for the case when the early time bin of Alice's and Bob's qubits both arrive after an even number of microwave $\pi$ pulses after its initialization. Thereafter we generalize this result and describe the practical consequences for the MDI-QKD protocol.

The sequence begins with a $\pi/2$ microwave pulse, preparing the spin in the state $\ket{\psi_i} = (\ket{\uparrow} + \ket{\downarrow})/\sqrt{2}$. In the absence of a photon at the device, the subsequent microwave $\pi$-pulses, which follow an XY8-N type pattern, decouple the spin from the environment and at the end of the sequence should preserve the spin state $\ket{\psi_i}$. 
However, reflection of Alice's photonic qubit $\ket{A} = (\ket{e} + e^{i \phi_1} \ket{l}) / \sqrt{2}$ from the device results in the entangled spin-photon state $\ket{\psi_A} = (\ket{\uparrow e} + e^{i\phi_1} \ket{\downarrow l})/\sqrt{2}$.
The full system is in the state 
\begin{equation}
	\label{eq:spinphoton}
	\ket{\psi_A} = \frac{\ket{+x}(\ket{\uparrow} + e^{i\phi_1}\ket{\downarrow}) + \ket{-x}(\ket{\uparrow} - e^{i\phi_1}\ket{\downarrow})}{2}.
\end{equation}

Regardless of the input photon state, there is equal probability to measure the reflected photon to be $\ket{\pm x}$.
Thus, measuring the photon in X basis (through the TDI) does not reveal the initial photon state.
After this measurement, the initial state of the photon $\ket{A}$ is teleported onto the spin: $\ket{\psi_{m_1}} = (\ket{\uparrow} + m_1 e^{i\phi_1} \ket{\downarrow})\sqrt{2}$, where $m_1=\pm1$ denotes the detection outcome of the TDI \cite{Nguyen2019, Kalb2015}.
The quantum state of Alice's photon is now stored in the spin state, which is preserved by the dynamical decoupling sequence.

Reflection of the second photon $\ket{B} = (\ket{e} + e^{i \phi_2} \ket{l})\sqrt{2}$ from Bob results in the spin-photon state $\ket{\psi_{m_1, B}} = (\ket{\uparrow e} + m_1 e^{i(\phi_1 + \phi_2)} \ket{\downarrow l})/\sqrt{2}$. 
This state 
now has a phase that depends on the initial states of both photons, enabling the photon-photon BSM measurements described below.
Rewriting Bob's reflected photon in the X basis,  the full system is in the state
\begin{multline}
	\label{eq:spinphoton}
	\ket{\psi_{m_1, B}}= \{\ket{+x}(\ket{\uparrow} + m_1 e^{i(\phi_1+\phi_2)}\ket{\downarrow}) \\ 
	+ \ket{-x}(\ket{\uparrow} - m_1 e^{i(\phi_1+\phi_2)}\ket{\downarrow})\}/2.
\end{multline}
The second measurement result $m_2$ once again contains no information about the initial state $\ket{B}$, yet heralds the final spin state $\ket{\psi_{m_1, m_2}} = (\ket{\uparrow} + m_1 m_2 e^{i(\phi_1 + \phi_2)} \ket{\downarrow})$ as described in the main text. 
When this state lies along the X axis of the Bloch sphere ($\phi_1 + \phi_2 = \{0, \pi\}$), the final result of the X basis measurement on the spin state $m_3$
has a deterministic outcome, dictated by all values of the parameters $\{\phi_1, \phi_2\}$ (known only to Alice and Bob) and $\{m_1, m_2\}$ (which are known to Charlie, but are completely random).
Conversely, all information available to Charlie $\{m_1, m_2, m_3\}$ only contains information on the correlation between the photonic qubits, not on their individual states. 
The resulting truth table for different input states is given in \etab{3}. For all input states, there is equal probability of measuring $\pm1$ for each individual measurement $m_i$. However, the overall parity of the three measurements $m_1 m_2 m_3$ depends on whether or not the input photons were the same, or opposite, for inputs $\ket{A}, \ket{B} \in \ket{\pm x}$ or $\ket{\pm y}$. 

We now address the fact that the BSM distinguishes either between $\{\ket{\Phi_\pm}\}$ or $\{\ket{\Psi_\pm}\}$ if there was an even or odd number of microwave $\pi$ pulses between incoming photons respectively.
This effect arises because
each $\pi$ pulse in the dynamical decoupling sequence toggles an effective frame change: $Y \leftrightarrow -Y$. 
The impact on this frame change on the BSM can be seen by writing the pairs of Bell states ($\ket{\Phi_\pm} = (\ket{ee} \pm \ket{ll})/\sqrt{2}$ and $\ket{\Psi_\pm} = (\ket{el} \pm \ket{le})/\sqrt{2}$) in the X and Y bases, where we have 
\begin{equation}
	\label{eq:Phipmx}
	\ket{\Phi_\pm}^{(X)} = (\ket{+x}\ket{\pm x} + \ket{\mp x}\ket{-x})/\sqrt{2} \\
\end{equation}
\begin{equation}
	\label{eq:Phipmy}
	\ket{\Phi_\pm}^{(Y)} = (\ket{+y}\ket{\mp y} + \ket{\pm y}\ket{-y})/\sqrt{2} \\
\end{equation}
\begin{equation}
	\label{eq:Psipmx}
	\ket{\Psi_\pm}^{(X)} = (\ket{+x}\ket{\pm x} - \ket{\mp x}\ket{-x})/\sqrt{2} \\
\end{equation}
\begin{equation}
	\label{eq:Psipmy}
	\ket{\Psi_\pm}^{(Y)} = i (\ket{+y}\ket{\pm y} - \ket{\mp y}\ket{-y})/\sqrt{2} \\
\end{equation}
For X basis inputs, as seen by Eq.~\ref{eq:Phipmx} and \ref{eq:Psipmx}, switching between $\{\ket{\Phi_\pm}\}$ and $\{\ket{\Psi_\pm}\}$ measurements does not affect the inferred correlation between input photons. For Y basis inputs however, this does result in an effective bit flip in the correlation outcome (see Eq.~\ref{eq:Phipmy} and \ref{eq:Psipmy}). In practice, Alice and Bob can keep track of each Y photon sent and apply a bit flip accordingly, as long as they have the appropriate timing information about MW pulses applied by Charlie. If Charlie does not give them the appropriate information, this will result in an increased QBER which can be detected.

\begin{figure}
\begin{center}
	\includegraphics[width=\linewidth]{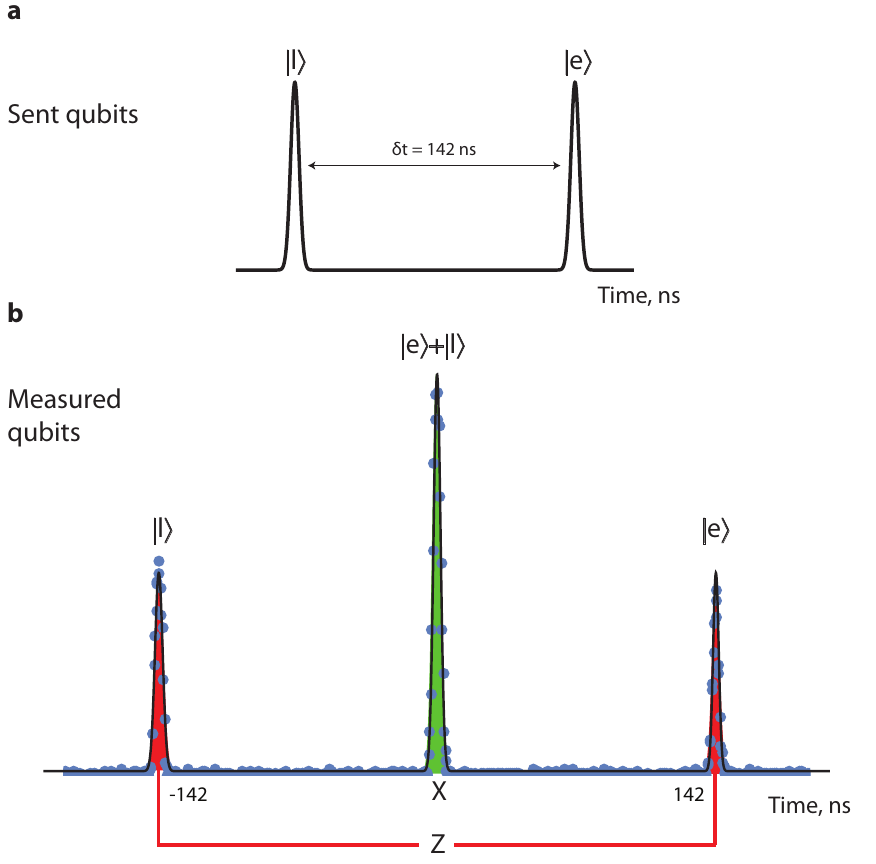}
\end{center}
\caption{{\bf Measurements on a single time-bin qubit in Z and X bases.}
		{\bf a,} Example of optical pulses sent for example in the experiment described in \fig{2d}.
		{\bf b,} Time trace of detected photons on + detector when pulses shown in (a) are sent directly into the TDI. The first and last peaks correspond to late and early photons taking the long and short paths of the TDI, which enable measurements in the Z basis $\{ \ket{e}, \ket{l} \}$. The central bin corresponds to the late and early components overlapping and interfering constructively to come out of the + port, equivalent to a measurement of the time bin qubit in the $\ket{+ x}$ state. A detection event in this same timing window on the - detector (not shown) would constitute a $\ket{-x}$ measurement.
		}
\end{figure}

As a final remark, this scheme also works for pairs of photons that are not both in the X or Y basis but still satisfy the condition $\phi_1 + \phi_2 = 0$. For example, $\ket{a}$ and $\ket{b}$ from \fig{3b} satisfy this condition. In this case, adequate correlations can still be inferred about the input photons, although they were sent in different bases. 

\begin{table}
	\begin{center}
		\begin{tabular}{ c | c || c | c}
			\hline
			Alice & Bob & Parity & Bell state \\
			\hline
			$\ket{+x}$ & $\ket{+x}$ & $+1$ & $\ket{\Phi_+}$ \\
			$\ket{+x}$ & $\ket{-x}$ & $-1$ & $\ket{\Phi_-}$ \\
			$\ket{-x}$ & $\ket{+x}$ & $-1$ & $\ket{\Phi_-}$ \\
			$\ket{-x}$ & $\ket{-x}$ & $+1$ & $\ket{\Phi_+}$ \\
			\hline
			$\ket{+y}$ & $\ket{+y}$ & $-1$ & $\ket{\Phi_-}$ \\
			$\ket{+y}$ & $\ket{-y}$ & $+1$ & $\ket{\Phi_+}$ \\
			$\ket{-y}$ & $\ket{+y}$ & $+1$ & $\ket{\Phi_+}$ \\
			$\ket{-y}$ & $\ket{-y}$ & $-1$ & $\ket{\Phi_-}$ \\
		\end{tabular}
		\caption{{\bf Truth table of asynchronous BSM protocol}, showing the parity (and BSM outcome) for each set of valid input states from Alice and Bob. In the case of Y basis inputs, Alice and Bob adjust the sign of their input state depending on whether it was commensurate with an even or odd numbered free-precession interval, based on timing information provided by Charlie.}
	\end{center}
\end{table}

\subsection{Test of Bell-CHSH inequality} In order to perform a test of the Bell-CHSH inequality \cite{Clauser1969}, we send input photons equally distributed from all states $\{\ket{\pm x}, \ket{\pm y }, \ket{\pm a}, \ket{\pm b}\}$ (\fig{3b}). We select for cases where two heralding events arise from input photons $\{A, B\} = \pm 1$ that are either $45^\circ$ or $135^\circ$ apart from one another. Conditioned on the parity outcome of the BSM ($\pm 1$), the Bell-CHSH inequality bounds the correlations between input photons as
\begin{equation}
	\label{eq:bell}
	S_\pm = |\braket{A \cdot B}_{xa} - \braket{A \cdot B}_{xb} - \braket{A \cdot B}_{ya} - \braket{A \cdot B}_{yb}| \leq 2,
\end{equation}
where the subscripts denote the bases the photons were sent in. The values of each individual term in Eq.~\ref{eq:bell}, denoted as ``input correlations," are plotted in \fig{3d} for positive and negative parity outcomes.

\section{Analysis of quantum communication experiment} 

\subsection{Estimation of QBER}
In order to achieve the lowest QBER, we routinely monitor the status trigger of the pre-selection routine and adjust the TDI. 
Additionally, we keep track of the timing when the TDI piezo voltage rails.
This guarantees that the SiV is always resonant with the photonic qubits and that the TDI performs high-fidelity measurements in X basis.
This is implemented in software with a response time of \SI{100}{\milli\second}. 

For each experiment, we estimate the QBER averaged over all relevant basis combinations. 
This is equivalent to the QBER when the random bit string has all bases occurring with the same probability, (an unbiased and independent basis choice by Alice and Bob). 
We first note that the QBER for positive and negative parity announcements are not independent.
We illustrate this for the example, that Alice and Bob send photons in the X basis. We denote the probability $P$ that Alice sent qubit $\ket \psi$, Bob sent qubit $\ket \xi$ and the outcome of Charlie's parity measurement is $m_C$, conditioned on the detection of a coincidence, as $P(\psi_A\cap \xi_B\cap m_C)$. We find for balanced inputs $P(+X_A\cap -X_B) = P(-X_A\cap +X_B)$ that $P(E_{XX}|+_C) = P(E_{XX}|-_C)$ with $E_{XX}$ denoting the occurrence of a bit error in the sifted key of Alice and Bob. We thus find for the posterior probability $L$ for the average QBER for XX coincidences
\begin{multline}
	L(P(E_{XX})) = L(P(-_C| +X_A\cap +X_B)) \\ 
	* L(P(+_C| +X_A\cap -X_B)) * L(P(+_C| -X_A\cap +X_B)) \\
	* L(P(-_C| -X_A\cap -X_B)).
\end{multline}
Note that this expression is independent of the actual distribution of $P(\psi_A\cap \xi_B)$. Here, the posterior probability $L(P(+_C| +X_A\cap -X_B))$ is based on the a binomial likelihood function $P(N_{m_C\cap \psi_A\cap \xi_B}|N_{\psi_A\cap \xi_B}, L)$, where $N_\mathcal{C}$ denotes the number of occurrences with condition $\mathcal{C}$. Finally the posterior probability of the unbiased QBER is $L(P(E)) = L(P(E_{XX})) * L(P(E_{YY}))$. All values presented in the text and figures are maximum likelihood values with bounds given by the confidence interval of $\pm34.1\%$ integrated posterior probability. Confidence levels towards a specific bound (for example, unconditional security \cite{Shor2000}) are given by the integrated posterior probability up to the bound.

\begin{figure*}
\begin{center}
	\includegraphics[width=\textwidth]{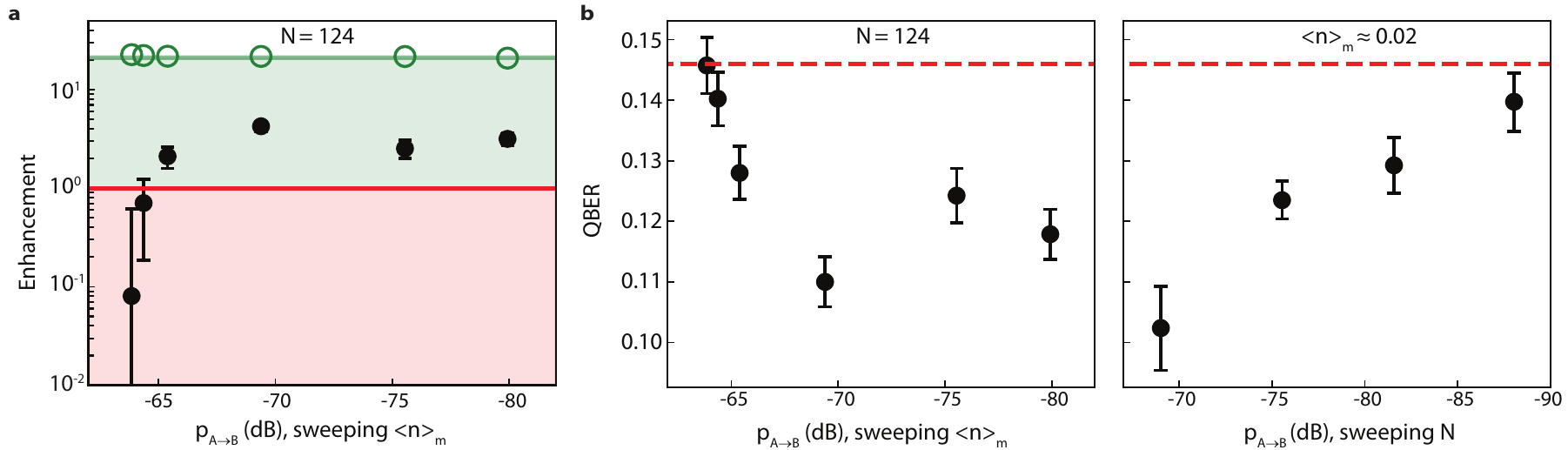}
\end{center}
\caption{{\bf Performance of memory-device versus of channel loss.}
		{\bf a,} Enhancement of memory-based approach compared to direct transmission approach, keeping $N = 124$ fixed and varying $\braket{n}_m$ in order to vary the effective channel transmission probability $p_{A\rightarrow B}$. At high $p_{A\rightarrow B}$ (larger $\braket{n}_m$), $r_s$ approaches $0$ due to increased QBER arising from undetected scattering of a third photon.
		{\bf b,} (Left) Plot of QBER for same sweep of $\braket{n}_m$ shown in {\bf a}.
		(Right) Plot of QBER while sweeping $N$ in order to vary loss. These points correspond to the same data shown in \fig{4}. At lower $p_{A\rightarrow B}$ (larger $N$), microwave-induced heating-related dephasing leads to increased QBER.
		}
\end{figure*}

To get the ratio of the distilled secret key rate with respect to the sifted key rate by (ideal) error correction and privacy amplification, we use the bounds given by difference in information by Alice and Bob with respect to a potential eavesdropper who performs individual attacks \cite{Gisin2002}: $r_s = I(A, B) - I (A/B, E)^{\mathrm{max}}$. We use the full posterior probability distribution of QBER (which accounts for statistical and systematic uncertainty in our estimate) to compute the error bar on $r_s$, and correspondingly, the error bars on the extracted secret key rates plotted in \fig{4}.

\subsection{Optimal parameters for asynchronous Bell state measurements}

We minimize the experimentally extracted QBER for the asynchronous BSM to optimize the performance of the memory node.
The first major factor contributing to QBER is the scattering of a third photon that is not detected, due to the finite heralding efficiency $\eta = 0.423\pm0.04$.
This is shown in \fig{2f}, where the fidelity of the spin-photon entangled state diminishes for $\braket{n}_m \gtrsim 0.02$.
At the same time, we would like to work at the maximum possible $\braket{n}_m$ in order to maximize the data rate to get enough statistics to extract QBER (and in the quantum communication setting, efficiently generate a key).

To increase the key generation rate per channel use, one can also fit many photonic qubits within each initialization of the memory.
In practice, there are 2 physical constraints: (1) the bandwidth of the SiV-photon interface and (2) the coherence time of the memory. 
We find that one can satisfy (1) at a bandwidth of roughly \SI{50}{\mega\hertz} with no measurable infidelity. 
For shorter optical pulses ($< 10$\,ns), the spin-photon gate fidelity is reduced. 
In principle, the SiV-photon bandwidth can be increased by reducing the atom-cavity detuning (here $\sim$ \SI{60}{\giga\hertz}) at the expense of having to operate at higher magnetic fields where microwave qubit manipulation is not as convenient \cite{Nguyen2019a}. 

\newcommand{\doublecell}[2][c]{%
	\begin{tabular}[#1]{@{}c@{}}#2\end{tabular}}

\begin{table*}
	\begin{center}
		\begin{tabular}{c | c | c | c | c}
			& \shortstack{per channel\\occupancy} & \shortstack{per channel\\occupancy} & \shortstack{per channel\\use} & \shortstack{per channel\\use}\\
			\hline
			X:Y basis bias & $50:50$ & $99:1$ & $50:50$ & $99:1$ \\
			\hline
			\hline
			Secure key rate $R$ $\left[10^{-7}\right]$& $1.19^{+0.14}_{-0.14}$ & $2.33^{+0.28}_{-0.28}$ & $2.37^{+0.29}_{-0.28}$ & $4.66^{+0.56}_{-0.55}$ \\
			\hline
			$ R / R_\textrm{max}(\textrm{X:Y})$ & $2.06^{+0.25}_{-0.25}$ &$2.06^{+0.25}_{-0.25}$ & $4.13^{+0.50}_{-0.49}$ & $4.13^{+0.50}_{-0.49}$ \\
			\hline
			$R/(1.44p_{A\rightarrow B})$ & $0.71^{+0.09}_{-0.08}$ & $1.40^{+0.17}_{-0.17}$ & $1.43^{+0.17}_{-0.17}$ & $2.80^{+0.34}_{-0.33}$\\
			$1-$confidence level & & $1.1^{+0.4}_{-0.3}\times 10^{-2}$ & $8^{+3}_{-2}\times 10^{-3}$ & $1.3^{+0.5}_{-0.3}\times 10^{-7}$
		\end{tabular}
		\caption{{\bf Quantum-memory-based advantage.} 
			Secret key rates with the asynchronous BSM device and comparison to ideal direct communication implementations, based on the performance of our network node for $N=124$ and $\braket{n}_m \sim 0.02$. Distillable key rates for $E=0.110 \pm 0.004$ for unbiased and biased basis choice are expressed in a per-channel-occupancy and per-channel-use normalization.
		Enhancement is calculated versus the linear optics MDI-QKD limit ($R_\textrm{max}(50:50)=p_{A\rightarrow B}/2$ for unbiased bases, $R_\textrm{max}(99:1) = 0.98 p_{A\rightarrow B}$ with biased bases) and versus the fundamental repeaterless channel capacity \cite{Pirandola2017} ($1.44p_{A\rightarrow B}$).
		Confidence levels for surpassing the latter bound \cite{Pirandola2017} are given in the final row.}
	\end{center}
\end{table*}

Even with just an XY8-1 decoupling sequence (number of $\pi$ pulses $N_\pi = 8$), the coherence time of the SiV is longer than $200\,\mu$s (\efig{3c}) and can be prolonged to the millisecond range with longer pulse sequences \cite{Nguyen2019}.
Unfortunately, to satisfy the bandwidth criteria (1) and to drive both hyperfine transitions (\efig{3a}), we must use short (\SI{32}{\nano\second} long $\pi$ pulses), which cause additional decoherence from ohmic heating \cite{Nguyen2019a} already at $N_\pi = 64$ (\efig{3e}).
Due to this we limit the pulse sequences to a maximum $N_\pi = 128$, and only use up to $\approx 20\,\mu$s of the memory time.
One solution would be to switch to superconducting microwave delivery.
Alternatively, one can use a larger value of $\tau$ to allow the device to cool down in between subsequent pulses \cite{Nguyen2019a} at the expense of having to stabilize a TDI of larger $\delta t$.
Working at larger $\delta t$ also enables temporal multiplexing by fitting multiple time-bin qubits per free-precession interval. In fact, with $2 \tau = $ \SI{142}{\nano\second}, even given constraint (1) and the finite $\pi$ time, we can already fit up to $4$ optical pulses per free-precession window, enabling a total number of photonic qubits of up to $N = 504$ for only $N_\pi = 128$.

In benchmarking the asynchronous BSM for quantum communication, we optimize the parameters $\braket{n}_m$ and $N$ to maximize our enhancement over the direct transmission approach, which is a combination of both increasing $N$ and reducing the QBER, since a large QBER results in a small secret key fraction $r_s$. As described in the main text, the effective loss can be associated with $\braket{n}_p$, which is the average number of photons per photonic qubit arriving at the device, and is given straightforwardly by $\braket{n}_p = \braket{n}_m / N$. The most straightforward way to sweep the loss is to keep the experimental sequence the same (fixed $N$) and vary the overall power, which changes $\braket{n}_m$. The results of such a sweep are shown in \efig{5a, b}. For larger $\braket{n}_m$ (corresponding to lower effective channel losses), the errors associated with scattering an additional photon reduce the performance of the memory device. 

Due to these considerations, we work at roughly $\braket{n}_m \lesssim 0.02$ for experiments in the main text shown in \fig{3 and 4}, below which the performance does not improve significantly. At this value, we obtain BSM successes at a rate of roughly \SI{0.1}{\hertz}. 
By fixing $\braket{n}_m$ and increasing N, we maintain a tolerable BSM success rate while increasing the effective channel loss.
Eventually, as demonstrated in \efig{5c} and in the high-loss data point in \fig{4}, effects associated with microwave heating result in errors that again diminish the performance of the memory node for large $N$. As such, we conclude that the optimal performance of our node occurs for $\braket{n}_m \sim 0.02$ and $N \approx 124$, corresponding to an effective channel loss of 
\SI{69}{dB} between Alice and Bob, which is equivalent to roughly \SI{350}{\kilo\meter} of telecommunications fiber.

We also find that the QBER and thus the performance of the communication link is limited by imperfect preparation of photonic qubits. Photonic qubits are defined by sending arbitrary phase patterns generated by the optical AWG to a phase modulator. For an example of such a pattern, see the blue curve in \fig{3a}. We use an imperfect pulse amplifier with finite bandwidth ($0.025-700$ \si{\mega\hertz}), and find that the DC component of these waveforms can result in error in photonic qubit preparation on the few $\%$ level. By using a tailored waveform of phases with smaller (or vanishing) DC component, we can reduce these errors. We run such an experiment during the test of the Bell-CHSH inequality. We find that by evaluating BSM correlations from $\ket{\pm a}$ and $\ket{\pm b}$ inputs during this measurement, we estimate a QBER of $0.097 \pm 0.006$. 

Finally, we obtain the effective clock-rate of the communication link by measuring the total number of photonic qubits sent over the course of an entire experiment. In practice, we record the number of channel uses, determined by the number of sync triggers recorded (see \efig{1a}) as well as the number of qubits per sync trigger ($N$). We then divide this number by the total experimental time from start to finish ($\sim$ 1-2 days for most experimental runs), including all experimental downtime used to stabilize the interferometer, readout and initialize the SiV, and compensate for spectral diffusion and ionization. 
For $N = 248$, we extract a clock rate of $\SI{1.2}{\mega\hertz}$. As the secret key rate in this configuration exceeds the conventional limit of $p/2$ by a factor of $3.8\pm1.1$, it is competitive with a standard MDI-QKD system operating at $4.5^{+1.3}_{-1.2}$ \si{\mega\hertz} clock rate.

\subsection{Performance of memory-assisted MDI-QKD}
A single optical link can provide many channels, for example, by making use of different frequency, polarization, or temporal modes. To account for this, when comparing different systems, data rates can be defined on a per-channel-use basis. 
In a MDI-QKD setting, full usage of the communication channel between Alice and Bob means that both links from Alice and Bob to Charlie are in use simultaneously. For an asynchronous sequential measurement, typically only half of the channel is used at a time, for example from Alice to Charlie or Bob to Charlie.
The other half can in principle be used for a different task when not in use. For example, the unused part of the channel could be routed to a secondary asynchronous BSM device. In our experiment, 
we can additionally define as a second normalization the rate per channel ``occupancy", which accounts for the fact that only half the channel is used at any given time.  
The rate per channel occupancy is therefore half the rate per full channel use. For comparison, we typically operate at $1.2\%$ channel use and $2.4\%$ channel occupancy.

To characterize the optimal performance of the asynchronous Bell state measurement device, we operate it in the optimal regime determined above ($N=124$, $\braket{n}_m \lesssim 0.02$). We note that the enhancement in the sifted key rate over direct transmission MDI-QKD is given by 
\begin{equation}
\frac{R}{R_\textrm{max}}=\eta^2 \frac{(N_\pi-1)(N_\pi-2)N_\textrm{sub}}{2N_\pi}
\end{equation}
and is independent of $\braket{n}_m$ for a fixed number of microwave pulses $N_\pi$ and optical pulses per microwave pulse $N_\textrm{sub}$ and thus fixed $N=N_\pi N_\textrm{sub}$. For low $\braket{n}_m$, three photon events become negligible and therefore QBER saturates, such that the enhancement in the secret key rate saturates as well (\efig{5a}). We can therefore combine all data sets with fixed $N=124$ below $\braket{n}_m \lesssim 0.02$ to characterize the average QBER of $0.116\pm0.002$ (\fig{3c}). The key rates cited in the main text relate to a data set in this series ($\braket{n}_m \approx 0.02$), with a QBER of $0.110\pm0.004$.  
A summary of key rates calculated on a per-channel use and per-channel occupancy basis, as well as comparisons of performance to ideal MDI-QKD and repeaterless bounds \cite{Pirandola2017} are given in \etab{4}.

Furthermore, we extrapolate the performance of our memory node to include biased input bases from Alice and Bob. This technique enables a reduction of channel uses where Alice and Bob send photons in different bases, but is still compatible with secure key distribution \cite{Lo2005}, allowing for enhanced secret key rates by at most a factor of $2$. The extrapolated performance of our node for a bias of 99:1 is also displayed in \etab{4}, as well as comparisons to the relevant bounds. We note that basis biasing does not affect the performance when comparing to the equivalent MDI-QKD experiment, which is limited by $p_{A\rightarrow B}/2$ in the unbiased case and $p_{A \rightarrow B}$ in the biased case. However, using biased input bases does make the performance of the memory-assisted approach more competitive with the fixed repeaterless bound \cite{Pirandola2017} of $1.44 p_{A \rightarrow B}$.

\end{document}